# A NOVEL CONTACT RESISTANCE MODEL OF ANISOTROPIC CONDUCTIVE FILM FOR FPD PACKAGING

*Gou-Jen Wang[1], Yi-Chin Lin[2] and Gwo-Sen Lin[3]*

[1,2] Department of Mechanical Engineering,
National Chung-Hsing University,
Taiwan

[3] Wintek Inc., Taiwan

**ABSTRACT**

In this research, a novel contact resistance model for the flat panel display (FPD) packaging based on the within layer parallel and between layers series resistance concepts was proposed. The FJ2530 anisotropic conductive films (ACF) by Sony Inc. containing the currently smallest 3μm conductive particles was used to conduct the experiments to verify the accuracy of the proposed model. Calculated resistance of the chip-on-glass (COG) packaging by the proposed model is 0.163Ω. It is found that the gold bump with 0.162Ω resistance play the major role of the overall resistance. Although the predicted resistance by the proposed model is only one third of the experimentally measured value, it has been three-fold improvement compared to the existing models.

## 1. INTRODUCTION

Resolution of a flat panel display (FPD) is determined by the channels of its controller IC. The definition of resolution is the number of pixels in each square inch of the FPD. For the liquid crystal display, each pixel is defined by the crossing of a horizontal scan line and a vertical source line which are connected to the controller IC. To increase the resolution of a FPD, the channels of the controller IC have to be increased. High resolution implies that dense and fine layout of the scan lines and source lines is required. Therefore more advanced techniques in packaging are desired to accurately connect the controller IC to the interface. The anisotropic conductive film (ACF) is the most commonly used conductive material in FPD packaging. In general, the conductive particles of the ACF distribute uniformly between the controller IC and the transparent electrode of the FPD after the packaging having been conducted. This leads to a high possibility of short circuit. To keep away the short circuit condition from happing, a distinct packaging mechanism from that of the general IC packaging is thus desired.

Moreover, the amount of ACF and the deformation of them during packaging determine the stability of the contact resistance of a FPD. A stable contact resistance always leads to a FDP with high quality. If a precise packaging model of the FPD is established such that the contact resistance of different structures can be accurately estimated, then the optimal design of any packaging structure can be easily achieved.

Huang [1] used the resistance model of particles between two arbitrary electrodes to estimate the total conductive resistance between two electrodes in FPD packaging. The resistance was found to be inversely proportional to the radius of the particle. In real world, the materials that contacts with the conducting particles are the gold bumps and the indium tin oxide (ITO) electrode. There is a big difference in electric conductivity between these two materials. The estimated conductive resistance based on the assumption of equal electric conductivity is not so realistic. Moreover, the inference that the total resistance is inversely proportional to the radius of the particle does not coincide with the Ohm's law that the resistance is inversely proportional to the area.

Yim et al. [2] simulated the theoretical electrical conduction model of the ACF in terms of physical contact mechanism. Two pressure dependent models, the elastic/plastic deformation and the finite element method, were implemented to investigate the changs of the contact area during bonding. Contributions to the equivalent resistance were assumed including the resistance of two electrodes, the particle-electrode contact resistance, the resistance across one particle, and the number of conduction paths. The proposed model was phenomenologically in good agreement with the measured results except at higher bonding pressure. However, the experimental resistance (about 0.6Ω) was still ten-fold of the predicted value (around 0.06Ω). The gap between the experiment and the prediction implies that except the compressive load, there are some unmodeled properties affecting the contact resistance.

Chin et al. [3] compared different models [2][4-6] having been used in estimating contact resistance of the anisotropic conductive adhesive assemblies. It was found that there are great inconsistencies among the current models and between the predicted contact resistances and the experimentally measured values. It was pointed out that tunneling resistance, multimaterial layers, edge effect, rough surfaces, elastic recovery and residual force, interaction between nearby particles and variation on the radii of multiple particles which are not included in existing models, are likely to be the causes of inconsistencies.

To improve the notable disagreement between the estimated contact resistances and the experimentally measured values of the existing models in anisotropic conductive adhesive assemblies, a novel model based on the within layer parallel and between layers series resistance concept is proposed in this article. The FJ2530 anisotropic conductive films (ACF) by Sony Inc. that contains the currently smallest 3μm conductive particles were used to conduct the experiments to verify the accuracy of the proposed model.

## 2. MODEL CONSTRUCTION

### 2.1 Equivalent Resistance

According to the mechanism of the chip-on-glass (COG) FPD packaging, the contact resistance is composed of resistances of (1) aluminum layer (2) barrier layer (3) gold bump (4) ACF layer (conductive particles) (5) ITO pad. The proposed model that assumes the contact resistance is the equivalent resistance of these five resistances in series connecting is schematically illustrated in Figure 1. In which, resistances of aluminum, the barrier layer, the gold bump, and the ITO pad can be calculated in terms of their sheet resistances and surface areas. The resistance of each conductive particle is approximated by the resistance of a cylinder that has similar electric conductivity as the conductive particle.

The equivalent contact resistance can be approximated as,

$$R_{eq} = R_{Al} + R_{UBM} + R_{Au\_bump} + R_{ACF} + R_{ITO} \quad (1)$$

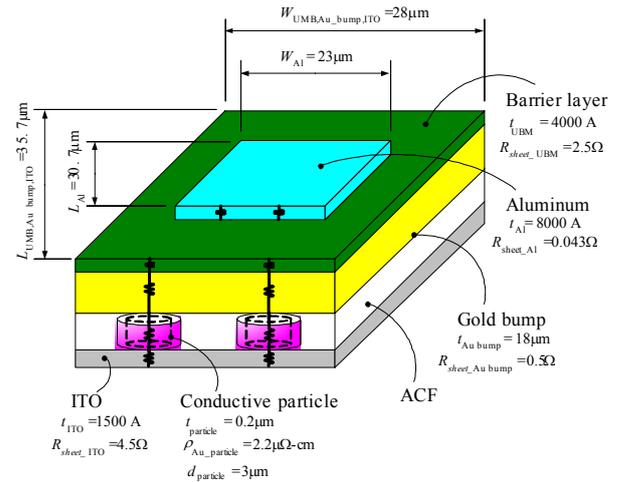

Figure 1. Schematic illustration of the proposed contact resistance model

### 2.2 Conversion of the sheet resistance

In general, the experimentally measured resistance is sheet resistance. It should be converted into the electric resistance in *z*-direction to enable the calculation of the equivalent resistance. The conversion details are as follows.

Assume the cuboid as shown in Figure 2 is conductive. Its resistance can be described by,

$$R = \rho \frac{l}{A} = \frac{\rho}{t}\frac{l}{w} = R_{sheet}\frac{l}{w} \quad (2)$$

where $R_{sheet}$ denotes the sheet resistance, defined as resistivity per unit length.

If $l=w$, then $R=R_{sheet}$. For a unit cube ($l=w=t$), resistances in all three directions are identical and equal to the sheet resistance. Therefore the resistance in *z*-axis of a conductive cube with volume $t^3$ can be represented by its sheet resistance.

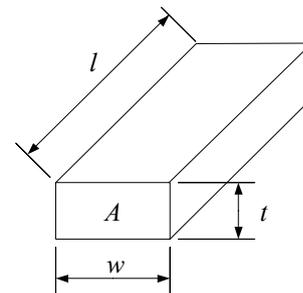

Figure 2. Sketch of a conductive cuboid

Assume a conductive layer is divided into *n* cubic units as shown in Figure 3. The equivalent resistance ($R_{layer}$) of this layer in *z*-axis is equal to the parallel connection resistance of these *n* cubic units ($n=WL/t^2$) as shown in Figure 4 and is calculated according to Equation (3).

$$\frac{1}{R_{layer}} = \frac{1}{R_1} + \frac{1}{R_2} + \ldots + \frac{1}{R_{n-1}} + \frac{1}{R_n} = n\frac{1}{R_{sheet}} \quad (3)$$

Therefore,

$$R_{layer} = R_{sheet}/n \quad (4)$$

Once the sheet resistance of a conductive layer has been measured, the resistance of this conductive layer in *z*-direction can be calculated by using the above equation.

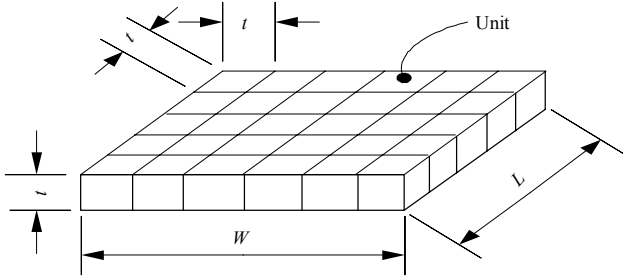

Figure 3. Conductive layer divided into *n* cubic units

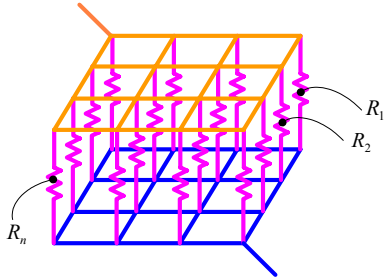

Figure 4. Equivalent circuit of the conductive layer in *z*-axis

**2.3 Calculation of the layer-by-layer resistance**

(1) Resistance of the aluminum layer

The sheet resistance of the aluminum layer provided by the vendor is $R_{sheet\_Al} = 0.043$ Ω. The number of the conductive cubes of the aluminum wire implemented in real experiment is $n_{Al} = WL/t^2 = 1103.28$. The equivalent resistance of the aluminum layer is then estimated to be,

$$R_{Al} = R_{sheet\_Al}/n = 0.043/1103.28 = 3.897 \times 10^{-5} \, \Omega$$

(2) Resistance of the barrier layer

The sheet resistance of the barrier layer provided by the vendor is $R_{sheet\_UBM} = 2.5$ Ω.

$$n_{UBM} = WL/t^2 = 6247.5$$

$$R_{UBM} = R_{sheet\_UBM}/n = 2.5/6247.5 = 4 \times 10^{-4} \, \Omega$$

(3) Resistance of the gold bump layer

The sheet resistance of the gold bump layer is $R_{sheet\_Au\,bump} = 0.5\,\Omega$.

$$n_{Au\,bump} = WL/t^2 = 3.085$$

$$R_{Au\,bump} = R_{sheet\_Au\,bump}/n = 0.5/3.085 = 0.162 \, \Omega$$

(4) Resistance of the ACF layer

After examining the cross sections of both the particle-ITO and particle-gold bump contact surfaces, it was found the particles deformed at the particle-ITO contact surface and stuck or deformed at the particle-gold bump contact surface. The effective contact surface between the conductive particle and others materials can be considered as that of an O-ring contact. For simplicity, the compressive load induced resistance is neglected and only the static contact resistance is considered. It is assumed that the contact resistance of each conductive particle is equivalent to that of the hollow cylinder as shown in Figure 5. The main reason for this assumption is that the statically conductive route connecting the gold bumps and the ITO electrode is similar to a cylinder conductor. Moreover, the metal thickness and the conductivity of the 3μm ACF used in experiments are presumed to be *t*=0.2μm and *ρ*=2.2μΩ-cm, respectively. The radius of the ACF is 1.5μm.

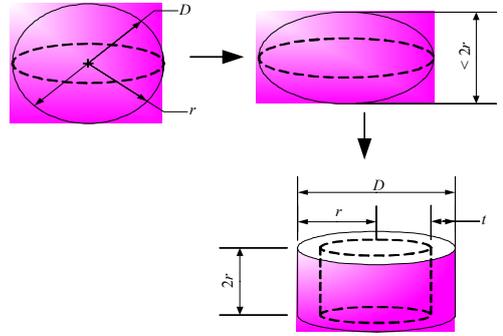

Figure 5. ACF contact surface approximation

The resistance of each conductive particle can be estimated by,

$$R_{single\_particle} = 2.2 \times 10^{-2} \frac{2 \times 1.5}{\pi \times (2.25 - 1.69)} = 3.75 \times 10^{-2} \, \Omega$$

The average number of the effectively contacting particles in area of 1500 μm² is 21.6. The overall resistance of the ACF layer is calculated by,

$$R_{ACF\_layer} = 3.75 \times 10^{-2}/21.6 = 0.0017 \, \Omega$$

(5) Resistance of the ITO pad

The sheet resistance of the ITO pad is $R_{sheet\_ITO} = 4.5$ Ω.

$$n_{ITO} = WL/t^2 = 11106.67$$

$$R_{ITO} = R_{sheet\_ITO}/n = 4.05 \times 10^{-4} \, \Omega$$

The equivalent resistance of the anisotropic conductive adhesive assembly is

$$R_{eq} = R_{Al} + R_{UBM} + R_{Au\_bump} + R_{AFC} + R_{ITO}$$
$$\approx 0.1645 \, \Omega$$

The major contribution to the equivalent resistance is the 0.162Ω from the gold bumps. Contributions from the rest layers are negligible.

## 2.4 Resistance vs. conductive particles

Figure 6 illustrates the dependence of the equivalent resistance on the number of conductive particles. For small number of particles, the resistance is inversely proportional to the amount of the particles. The resistance reaches a stable value of 1.63Ω when more than five particles are implemented.

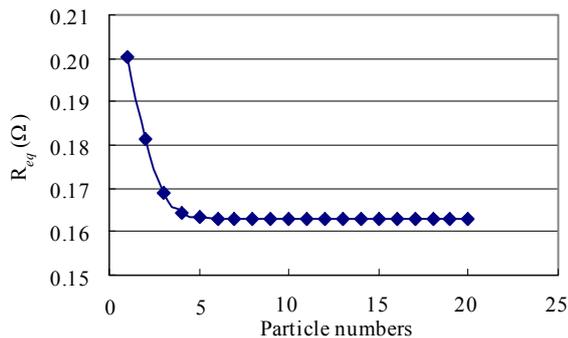

Figure 6. Dependency of resistance on particle numbers

## 2.5 Resistance vs. metal thickness of particle

Figure 7 describes the influence of the metal thickness of particle on the resistance. Similarly to the result of the number of particles, the resistance is inversely proportional to the metal thickness of particle.

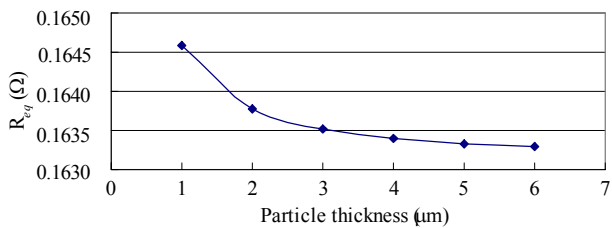

Figure 7. Dependency of resistance on metal thickness of particle

## 3. EXPERIMENTAL VERIFICATION

The currently smallest 3μm conductive particles made by Sony Inc. were used to conduct the COG experiments to verify the accuracy of the proposed model.

### 3.1 Experimental apparatus and devices
(1)Four-point resistivity measurement

To determine the resistivity of the 3μm ACF, the four point probe method [8] was implemented. The key advantage of the four point probe method is that the resistivity of the desired route can be accurately measured, without being affected by others routes. In our applications, the measurement noise due to the ITO electrode can be effectively excluded.

(2) Design of IC mask

The pattern of the IC mask shown in Figure 8 is composed of 18 test element groups (TEG). Each TEG contains (*i*)TEG01: formed by 4 gold bumps with 3 of them being shorted circuit to examine the point contact resistance; (*ii*) TEG02: formed by 4 gold bumps with 2 of them being shorted circuit to investigate the resistance of individual insulating film; (*iii*)TEG03: formed by 8 separated gold bumps to observe the impressive load induced short circuit problem and the uniformity of pressure, ensuring the reliability of the inspecting points.

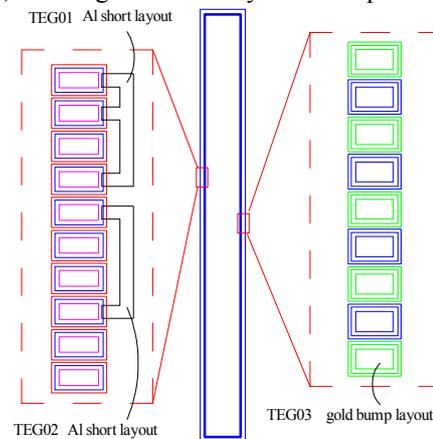

Figure 8. Layout of the IC test area

(3)Design of ITO mask

The mask for ITO layout is a 300mm x 400mm chromium mask. Layout of the ITO is required to match that of the IC TEG01~03 to enable effective measurement of resistance. Therefore, the TEG layout of the ITO mask includes (1)TEG01: formed by 4 ITO routes, in which routes #3 and #4 were shorted (Figure 9); (2)TEG02: formed by 2 independent ITO routes to successfully examine the resistance of the insulating film; (3)TEG03: formed by 8 ITO routes with the odd numbers routes being shorted and the even number routes being shorted (Figure 10).

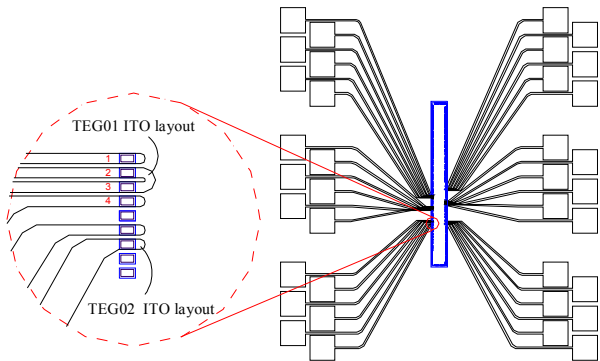

Figure 9. Layout of the glass substrate

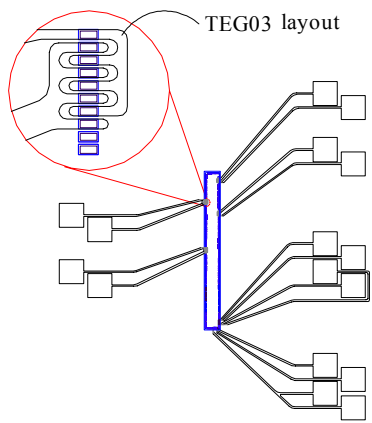

Figure 10. Layout of the multi insulating routes

(4) Anisotropic conductive film

The ACF implemented in this research is the FJ2530 by Sony Inc. with 25μm in thickness and 3.0mm in width. The conductive particles inside the ACF are the currently smallest 3μm particles.

(5) Glass substrate

The glass/$SiO_2$/ITO substrate made by Wintek Inc. was selected to be the glass substrate. Properties of the ITO substrate include: dimension=300mm x 400mm, ITO film thickness = 4500Å, sheet resistance = 4.5Ω/square.

### 3.2 Results and discussions

Recall Figure 6. The stable contact resistance of a chip-on-glass packaging predicted by the proposed model is between 0.16Ω and 0.2Ω. Experimentally measured value by implementing the 3μm conductive particles is around 0.5Ω. The ratio of the measured resistance to the predicted value is 3.04.

Comparing the result with the ratio of 10 obtained by other models [2][4-6], the proposed scheme predicts much more accurately. The main reason for the improvement is presumed that the route of gold bumps which contributes the most to the equivalent resistance is taken into consideration.

### 4. CONCLUSIONS

The equivalent contact resistance of a chip on glass (COG) FPD packaging is composed of resistances of (1) aluminum layer (2) barrier layer (3) gold bump (4) ACF layer (conductive particles) (5) ITO pad. In this research, a novel contact resistance model for the flat panel display (FPD) packaging based on the within layer parallel and between layers series concept was proposed. The FJ2530 ACF by Sony Inc. containing the currently smallest 3μm conductive particles was implemented to experimentally verify the accuracy of the proposed model.

Stable contact resistance predicted by the proposed model is 0.163Ω. Experimentally measured value by implementing the 3μm conductive particles is around 0.5Ω. The ratio of the measured resistance to the predicted value is 3.04. Comparing the results with that obtained by other models, it has been three-fold improvement.